\documentclass[prb,showpacs,twocolumn,superscriptaddress]{revtex4-1}
\usepackage{amssymb}
\usepackage{graphicx}
\usepackage{amsmath,amssymb}
\begin{document}

\title{Kondo effect of an adatom in graphene and its scanning tunneling
spectroscopy}
\author{Lin Li}
\affiliation{Center of Interdisciplinary Studies and Key Laboratory for Magnetism and
Magnetic Materials of the Ministry of Education, Lanzhou University, Lanzhou
730000, China}
\author{Yang-Yang Ni}
\affiliation{Center of Interdisciplinary Studies and Key Laboratory for Magnetism and
Magnetic Materials of the Ministry of Education, Lanzhou University, Lanzhou
730000, China}
\author{Yin Zhong}
\affiliation{Center of Interdisciplinary Studies and Key Laboratory for Magnetism and
Magnetic Materials of the Ministry of Education, Lanzhou University, Lanzhou
730000, China}
\author{Tie-Feng Fang}
\affiliation{Center of Interdisciplinary Studies and Key Laboratory for Magnetism and
Magnetic Materials of the Ministry of Education, Lanzhou University, Lanzhou
730000, China}
\author{Hong-Gang Luo}
\affiliation{Center of Interdisciplinary Studies and Key Laboratory for Magnetism and
Magnetic Materials of the Ministry of Education, Lanzhou University, Lanzhou
730000, China}
\affiliation{Beijing Computational Science Research Center, Beijing 100084, China}
\date{\today }


\begin{abstract}
We study the Kondo effect of a single magnetic adatom on the surface of
graphene. It was shown that the unique linear dispersion relation near the
Dirac points in graphene makes it more easy to form the local magnetic
moment, which simply means that the Kondo resonance can be observed in a
more wider parameter region than in the metallic host. The result indicates
that the Kondo resonance indeed can form ranged from the Kondo regime, to
the mixed valence, even to the empty orbital regime. While the Kondo resonance
displays as a sharp peak in the first regime, it has a peak-dip structure and/or an
anti-resonance in the remaining two regimes, which result from the Fano
resonance due to the significant background leaded by dramatically
broadening of the impurity level in graphene. We also study the scanning
tunneling microscopy (STM) spectra of the adatom and they show obvious
particle-hole asymmetry when the chemical potential is tuned by the gate
voltages applied to the graphene. Finally, we explore the influence of the
direct tunneling channel between the STM tip and the graphene on the Kondo
resonance and find that the lineshape of the Kondo resonance is unaffected,
which can be attributed to unusual large asymmetry factor in graphene. Our
study indicates that the graphene is an ideal platform to study
systematically the Kondo physics and these results are useful to further
stimulate the relevant experimental studies on the system.
\end{abstract}

\maketitle

\section{Introduction} \label{sec1}

The experimental realization of graphene \cite{Novoselov2004, Novoselov2005}
composed of a monolayer of carbon atoms trigged a new wave to study the
carbon-based materials, both from fundamental physics and applications.
\cite{Beenakker2008, Neto2009, Peres2010} The graphene possesses perfect
two-dimensional massless Dirac fermion behaviors, and the valence and
conduction bands touch at two inequivalent Dirac points $K_{-}$ and $K_{+}$
at the corner of the Brillouin zone. Around the Dirac points the low-energy
excitations are linear and the unique electronic structure leads to a number
of unusual electronic correlation and transport properties of graphene.
\cite{Beenakker2008, Neto2009, Peres2010, Peres2006, Ziegler2006, Nomura2007,
Zhuang2009, Kotov2010, Jacob2010}

For example, the localized magnetic moment on adatom with inner shell
electrons in graphene behaves quite different to that in metals, a
conventional Fermi liquid with a constant density of states near the Fermi
surface. Anderson showed that in such a host the strongly interacting
impurity ion can be in a magnetic state if the on-site interaction $U$ is
sufficiently large and/or the hybridization $V$ between the impurity ion and
the conduction electrons is sufficiently small. \cite{Anderson1961}
Explicitly, it is required that the energy of the single occupancy states $%
\varepsilon_d$ is below the Fermi level $\mu$ but the energy of the double
occupancy states, namely, $\varepsilon_d + U$, should be larger than $\mu$.
Moreover, the magnetic regime was found to be symmetric around $y = (\mu -
\varepsilon_d)/U = 0.5$. For details one can refer to the phase diagram
given by the well-known Ref. [\onlinecite{Anderson1961}]. In contrast, when
the host is graphene, the situation is dramatically different due to the
linear density of states near the Dirac points. \cite{Uchoa2008,
Cornaglia2009, Hu2011} Firstly, the magnetic phase diagram is not symmetric,
it depends on that the impurity level $\varepsilon_d$ is above or below the
Dirac points. Secondly and also most important difference is that the phase
boundary is not symmetric around $y = 0.5$ in both cases. When $%
\varepsilon_d $ is above the Dirac points, the magnetic state extends to $y
< 0$, which means that even the impurity level is located above the Fermi
energy, it is possible that the localized magnetic moment exists. This is
quite different to that in usual metallic host. In the other case that $%
\varepsilon_d$ is below the Dirac points, the phase boundary extends to $y >
1$, which means that even the impurity is in the double occupancy state, it
can be magnetized. It was realized that all these unusual features are due
to the linear dispersion of the graphene around the Dirac points, and the
impurity level is broadened dramatically by hybridization to cross the Fermi
level. \cite{Uchoa2008}

A direct consequence of the much wide magnetic regime is that the Kondo
effect in graphene \cite{Sengupta2008} may exist in a more wide parameter
regime than that in the usual metallic host in which the Kondo effect only
exists in the Kondo regime, namely, $\epsilon_d/\Gamma \ll -0.5$, \cite%
{Goldhaber-Gordon1998} where $\Gamma$ is the effective tunneling coupling
between the impurity and conduction electrons. Moreover, it is also interesting to explore what effect the dramatic broadening of the impurity level on the Kondo physics. This motivates us to study the Kondo effect of an adatom on the
surface of the graphene sheet in a systematic way.

\begin{figure}[t]
\center{\includegraphics[clip=true,width=\columnwidth]{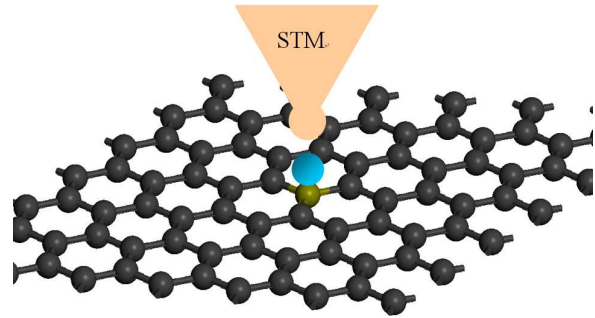}}
\caption{Schematic diagram of the STM measurement of a magnetic adatom
absorbed on the top of a carbon atom site. }
\label{fig1}
\end{figure}

In this work the system we are interested in consists of the adatom, the graphene,
and the tip on the top of the adatom, as shown in Fig. \ref{fig1}. As the
first step, we explore the intrinsic equilibrium Kondo effect of the adatom
as the conduction sea of electrons is limited to the graphene. The Kondo
physics of the adatom generally depends on its explicit positions on the
surface of the graphene. When the adatom is absorbed on the top of carbon
atom, the situation is quite simple since the orbital degree of freedom is
not involved in the case. However, when the adatom is located at the center
of the honeycomb, a high symmetry position, the adatom hybridizes with the
carbon atoms in both sublattices, \cite{Uchoa2009, Saha2010, Wehling2010a,
Wehling2010b, Jacob2010, Uchoa2011} thus the orbital degree of freedom is
involved which complicates the Kondo physics. For simplify, here we only
focus on the first case and leave the latter one for the future study. When
the Fermi energy is near the Dirac points, the Kondo effect is absent since a critical coupling strength must be satisfied for developing Kondo effect. \cite{Sengupta2008, Zhuang2009, Vojta2010} Once the Fermi energy is tuned to be
away from the Dirac points, despite it is below or above the Dirac points,
the Kondo resonance is found to exist in a wide parameter regime ranged from
the Kondo regime to the mixed valence regime, even to the empty orbital
regime. This is consistent with our intuition since the localized magnetic
moment exists in a very wide parameter regime. When the system is in the
Kondo regime, the Kondo resonance manifests as a sharp peak. This peak
appears weaker and weaker when the impurity level becomes deeper and deeper.
This is understandable since the broadening of the impurity level becomes
weak due to the low density of states of the graphene near the Dirac points.
When the system is in the mixed valence or empty orbital regimes, the Kondo
resonance develops, however, it does not show a sharp peak but a peak-dip
structure, even an anti-resonance, which is a typical Fano resonance
behavior. \cite{Fano1961, Luo2004}

In the next step, we consider a normal tip on the top of the adatom and the
aim is to study the scanning tunneling microscopy (STM) spectra of the Kondo
resonance of the adatom in graphene. We first turn off the direct tunneling
between the tip and the graphene. The Kondo resonance shows asymmetric
behaviors that depends on the position of the Fermi energy, namely, it is
above or below the Dirac points. Moreover, the Kondo temperature strongly
but asymmetrically depends on the explicit position of the Fermi energy
which is in agreement with reported in the literature. \cite{Vojta2010,Zhu2011}
Subsequently, we turn on the direct tunneling between the tip and the
graphene to study its influence on the lineshape of the Kondo resonance,
which also is more realistic case. It is somehow surprising that the
lineshape is almost unaffect by the additional direct channel. This result in fact
originates from the unusual large asymmetry factor in graphene.

The paper is organized as follows. In Sec. \ref{sec2} we take the
single-impurity Anderson model to describe the adatom on the surface of the
graphene and derive transport formula by Keldysh nonequilibrium Green's
function theory. In Sec. \ref{sec3} we solve numerically the equations
obtained in a self-consistent way and discuss explicitly the Kondo effect in
various situations including the cases of (i) the adatom plus the graphene,
(ii) the adatom plus the graphene plus the tip, and (iii) the STM spectra
with the direct channel between the tip and the graphene. Finally, a brief
summary is devoted to Sec. \ref{sec4}.

\section{Model and Formalism}

\label{sec2}

The Hamiltonian of a magnetic adatom on graphene with an STM tip consists of
the following terms
\begin{equation}
H=H_{d}+H_{g}+H_{dg}+H_{t}+H_{dt}+H_{tg},  \label{H}
\end{equation}
where
\begin{equation}
H_{d}=\sum_{\sigma }\varepsilon _{d}d_{\sigma }^{\dag }d_{\sigma }+U \,
d_{\sigma }^{\dag }d_{\sigma }d_{\bar{\sigma}}^{\dag }d_{\bar{\sigma}},
\label{Hd}
\end{equation}
describes the magnetic adatom with the impurity level $\varepsilon_d$ and
the on-site interaction $U$. \cite{Anderson1961} Here $d_{\sigma }^{\dag
}\left( d_{\sigma }\right) $ is a creation (annihilation) operator of
localized electrons on magnetic adatom with spin $\sigma \left( =\uparrow
,\downarrow \right) $ and $\bar\sigma = -\sigma$. The second term in Eq. (%
\ref{H}) is the tight-banding Hamiltonian describing graphene, which can be
written as in momentum space
\begin{equation}
H_{g}=-t\sum_{k\sigma }\left( \eta _{k}a_{k\sigma }^{\dagger }b_{k\sigma
}+H.c.\right) ,  \label{Hg}
\end{equation}
$a_{k\sigma }^{\dagger }\left( a_{k\sigma }\right) $ and $b_{k\sigma
}^{\dagger }\left( b_{k\sigma }\right) $ are creation (annihilation)
operators of electrons\ in sublattice A and B, respectively. $t$ is the
hopping energy between the nearest-neighbor carbon atoms, $\eta
_{k}=\sum_{_{i=1}}^{3}e^{i\vec{k}\cdot \vec{r}_{i}}\ $ with $\vec{r}_{1}=a%
\vec{x},$ $\vec{r}_{2}=-\frac{a}{2}\vec{x}+\frac{\sqrt{3}a}{2}\vec{y}$, $%
\vec{r} _{3}=-\frac{a}{2}\vec{x}-\frac{\sqrt{3}a}{2}\vec{y}$, and $a\approx
1.42$ \AA\ is the carbon-carbon spacing. \cite{Neto2009} The low-energy
transport properties of graphene are mostly determined by the nature of
spectrum around Dirac points $K_{\pm }$ at the corners of Brillouin zone,
where the dispersion is $\varepsilon _{k}\approx \pm \upsilon _{F}|k|$, the
sign $+\left( -\right) $ corresponds to the conduction (valence) band, and $%
\upsilon _{F}=3ta/2$ ($\approx 10^{6}$ m/s) is the Fermi velocity of Dirac
electrons. The tip Hamiltonian is described by a normal metal
\begin{equation}
H_{t}=\sum_{k\sigma}\xi _{k}t_{k\sigma}^{\dag}t_{k\sigma},  \label{Ht}
\end{equation}
where $t_{k\sigma}^{\dag}\left( t_{k\sigma}\right) $ represents the creation
(annihilation) of an electron with energy $\xi _{k}$ on tip. The Hamiltonian
\begin{equation}
H_{dg}=\sum_{k\sigma }\left( V_{ka\sigma }a_{k\sigma }^{\dagger
}+V_{kb\sigma }b_{k\sigma }^{\dagger }\right) d_{\sigma }+H.c.,  \label{Hdg}
\end{equation}
and
\begin{equation}
H_{dt}=\sum_{k\sigma }\left( V_{kt\sigma }t_{k\sigma }^{\dag }d_{\sigma
}+H.c.\right),  \label{Hdt}
\end{equation}
represent electrons tunneling from magnetic adatom to graphene and to STM
tip, respectively. $V_{ka\sigma }$ $\left( V_{kb\sigma }\right)$ is the
electron tunneling amplitude between magnetic adatom and sublattice A (B),
and $V_{kt\sigma }$ is the tunneling amplitude between magnetic adatom and
STM tip. If the adatom is adsorbed above the carbon atom in the sublattice
A, thus $V_{ka\sigma}=V_{kg\sigma }$ and $V_{kb\sigma }=0$ or in the
sublattice B, thus $V_{ka\sigma}=0 $ and $V_{kb\sigma }=V_{kg\sigma}$. The
direct tunneling Hamiltonian between graphene and the STM tip is described
by
\begin{equation}
H_{tg}=\sum_{kk^{\prime }\sigma }T_{tg}\left( t_{k\sigma }^{\dagger
}b_{k^{\prime }\sigma }+t_{k\sigma }^{\dagger }a_{k^{\prime }\sigma }\right)
+H.c.,  \label{Htg}
\end{equation}
where $T_{tg}$ is the hopping amplitude between the STM tip and graphene
sheet.

In steady state, the current from the STM tip to the graphene sheet can be
calculated by time-dependent occupancy number
\begin{eqnarray}
I && = -\frac{e}{2}\left( \left\langle \dot{N}_{t}\right\rangle
-\left\langle \dot{N} _{g}\right\rangle \right) = -\frac{ie}{2\hbar }%
\left\langle \left[ H,\left( N_{t}-N_{g}\right) \right] \right\rangle  \notag
\\
&& = -\frac{e}{\hbar }\text{Re}\sum_{k\sigma }\left( V_{kg\sigma }^{\ast
}G_{\sigma ,gk\sigma }^{<}\left( t,t^{\prime }\right) -V_{kt\sigma }^{\ast
}G_{\sigma ,tk\sigma }^{<}\left( t,t^{\prime }\right) \right)  \notag \\
&&\hspace{1cm}-\frac{2e}{\hbar }\text{Re}\sum_{kk^{\prime }\sigma
}T_{tg}^{\ast }G_{tk\sigma ,gk^{\prime }\sigma }^{<}\left( t,t^{\prime
}\right) ,  \label{I-N}
\end{eqnarray}%
where $N_{t}=\sum_{k\sigma }t_{k\sigma }^{\dag }t_{k\sigma }$ and $%
N_{g}=\sum_{k\sigma }c_{k\sigma }^{\dag }c_{k\sigma }$. Here we use $%
c_{k\sigma }^{\dagger }$($c_{k\sigma }$) to denote the electron creation
(annihilation) operator in the graphene sheet. The current is related to the
lesser Green's functions defined as $G_{\sigma,tk\sigma }^{<}\left(
t,t^{\prime }\right) =i\left\langle t_{k\sigma }^{\dag }\left( t^{\prime
}\right) d_{\sigma }\left( t\right) \right\rangle ,$ $G_{\sigma ,gk\sigma
}^{<}\left( t,t^{\prime }\right) =i\left\langle c_{k\sigma }^{\dagger
}\left( t^{\prime }\right) d_{\sigma }\left( t\right) \right\rangle $ and $%
G_{tk\sigma ,gk^{\prime }\sigma }^{<}\left( t,t^{\prime }\right)
=i\left\langle c_{k^{\prime }\sigma }^{\dagger }\left( t^{\prime }\right)
t_{k\sigma }\left( t\right) \right\rangle $. These Green's functions can be
calculated by the equations of motion (EOM) approach based on nonequilibrium
Green's functions theory. For a detail one can refer to Ref. [%
\onlinecite{Jauho1994}]. For simplify, we assume that the tunneling
amplitudes $V_{kg\sigma }=V_{g}$ and $V_{kt\sigma }=V_{t}$ are independent
on energy and spin. After some straightforward derivations, one can obtain the
current formula
\begin{widetext}
\begin{eqnarray}
I && =\frac{ie}{h}\sum_{\sigma }\int \left\{ \left( G_{\sigma }^{r}\left(
\varepsilon \right) -G_{\sigma }^{a}\left( \varepsilon \right) \right)
[\Gamma _{t}\left( \varepsilon \right) f_{t}\left( \varepsilon \right)
-\Gamma _{g}\left( \varepsilon \right) f_{g}\left( \varepsilon \right)
]+\left( \Gamma _{t}\left( \varepsilon \right) -\Gamma _{g}\left(
\varepsilon \right) \right) G_{\sigma }^{<}\left( \varepsilon \right)
\right\} d\varepsilon   \notag \\
&&\hspace{1cm} +\frac{2e}{h}\int \left( f_{t}\left( \varepsilon \right) -f_{g}\left(
\varepsilon \right) \right) \left[ \Pi \left( \varepsilon \right)
- \Phi \left(\varepsilon \right)\sum_{\sigma }\text{Im}(G_{\sigma }^{r}\left( \varepsilon \right) )\right] d\varepsilon, \label{I}
\end{eqnarray}
\end{widetext}where $G_{\sigma }^{<, r, a }\left( \varepsilon
\right) $ is the Fourier transformation of time-order Green's function $
G_{\sigma }^{<, r, a} \left( t,t^{\prime }\right) $ defined by $%
G_{\sigma }^{<}\left( t,t^{\prime }\right) = i\left\langle d_{\sigma }^{\dag
}\left( t^{\prime }\right) d_{\sigma }\left( t\right) \right\rangle$,  $
G_{\sigma }^{r\left( a\right) }\left( t,t^{\prime }\right) =\mp i\theta
\left( \pm t\mp t^{\prime }\right) \left\langle \{ d_{\sigma }\left(
t\right) ,d_{\sigma }^{\dag }\left( t^{\prime }\right) \}\right\rangle.$ $\Gamma _{g}\left( \varepsilon \right) =\pi \left\vert
V_{g}\right\vert ^{2}\rho _{g}\left( \varepsilon \right) $ and $\Gamma
_{t}\left( \varepsilon \right) =\pi \left\vert V_{t}\right\vert ^{2}\rho
_{t}\left( \varepsilon \right)$ are the couplings between the magnetic
adatom to the graphene and to the tip, respectively. $\Pi \left( \varepsilon
\right) =\pi \left\vert T_{tg}\right\vert ^{2}\rho _{t}\left( \varepsilon
\right) \rho _{g}\left( \varepsilon \right) $ corresponds to hopping from
the tip to graphene and vice versa. $\Phi \left( \varepsilon \right) =\sqrt{%
\pi \Pi \left( \varepsilon \right) \Gamma _{g}\left( \varepsilon \right)
\Gamma _{t}\left( \varepsilon \right) }$ is related to the interference
between the different tunneling channels. $\rho _{g}\left( \varepsilon
\right) $ and $\rho _{t}\left( \varepsilon \right) $ are the density of
states of graphene and STM tip, respectively. The density of states for tip
is simplified with a constant $\rho _{t}\left( \varepsilon \right) =1/2D$,
and the density of states for graphene can be evaluated with $\rho
_{g}\left( \varepsilon \right) =\Theta (D-|\varepsilon |)|\varepsilon
|/D^{2} $. \cite{Wehling2010b} Here $D$ is the half-width of conduction band. For
convenience, we take $\Gamma _{g}(\varepsilon )=\Gamma _{0}|\varepsilon |/{D}
$ where $\Gamma _{0}=\pi |V_{g}|^{2}/D$. $f_{g}\left( \varepsilon \right) $
and $f_{t}\left( \varepsilon \right) $ are the Fermi distribution functions
of graphene and the STM tip. The nonequilibrium Kondo effect can be
discussed by differential conductance near the Fermi level. Thus, our final
task to calculate the current is to solve the Green's functions involved.

The equations for the retarded Green's functions and the lesser Green's
functions can be derived by the Schwinger-Keldysh perturbation formalism
\cite{Mahan1993, Niu1999, Swirkowicz2003, Krawiec2004} which are given by,
respectively,
\begin{eqnarray}
&&G_{\sigma }^{r}\left( \varepsilon \right) =g_{d}^{r}(\varepsilon )\left(
1+\left\langle \left\langle \left[ d_{\sigma },H_{I}\right] ;d_{\sigma
}^{\dagger }\right\rangle \right\rangle _{\varepsilon }^{r}\right) ,
\label{E-1} \\
&&G_{\sigma }^{<}\left( \varepsilon \right) =g_{d}^{<}(\varepsilon )\left(
1+\left\langle \left\langle \left[ d_{\sigma },H_{I}\right] ;d_{\sigma
}^{\dagger }\right\rangle \right\rangle _{\varepsilon }^{a}\right)  \notag \\
&&\hspace{2cm}+g_{d}^{r}\left( \varepsilon \right) \left\langle \left\langle %
\left[ d_{\sigma },H_{I}\right] ;d_{\sigma }^{\dagger }\right\rangle
\right\rangle _{\varepsilon }^{<},  \label{E-2}
\end{eqnarray}%
where $g_{d}^{r}(\varepsilon )=1/\left( \varepsilon -\varepsilon
_{d}+i0^{+}\right) $ and $g_{d}^{<}\left( \varepsilon \right) =i\pi
f(\varepsilon )\delta (\varepsilon -\varepsilon _{d})$ are the
noninteracting retarded and lesser Green's functions for the isolated
magnetic adatom. In the above expressions the $H_{I}=Ud_{\sigma }^{\dag
}d_{\sigma }d_{\bar{\sigma}}^{\dag }d_{\bar{\sigma}}+H_{dg}+H_{dt}$ includes
the on-site Coulomb repulsion term on the adatom and the tunneling terms.
While the first term will lead to higher-order Green's functions contained
into $\left\langle \left\langle \left[ d_{\sigma },H_{I}\right] ;d_{\sigma
}^{\dagger }\right\rangle \right\rangle _{\varepsilon }^{r(a,<)}$ by using
Zubarev notations, \cite{Zubarev1960} the remaining two terms only lead to
the hybridization effect. Furthermore, the derivation of the EOM of the
higher-order Green's function will also introduce more higher-order Green's
functions, which have to be truncated in order to close the hierarchy of the
EOM of the successive Green's functions. Here we truncate this hierarchy of
EOM by using Lacroix's scheme, \cite{Lacroix1981} which is shown to be
enough to capture the Kondo resonance, even at zero temperature. \cite%
{Lacroix1981, Luo1999} Since the derivation is standard we directly write
down the Green's functions in the infinite-U limit as follows
\begin{equation}
G_{\sigma }^{r}\left( \varepsilon \right) =\frac{1-\left\langle n_{\bar{%
\sigma}}\right\rangle -A_{\bar{\sigma}}\left( \varepsilon \right) }{%
\varepsilon -\varepsilon _{d}-\Sigma \left( \varepsilon \right) \left( 1-A_{%
\bar{\sigma}}\left( \varepsilon \right) \right) -B_{\bar{\sigma}}\left(
\varepsilon \right) },  \label{G-1}
\end{equation}%
and
\begin{widetext}
\begin{equation}
G_{\sigma }^{<}\left( \varepsilon \right) =G_{\sigma }^{r}\left( \varepsilon
\right) \frac{\Sigma ^{<}\left( \varepsilon \right) \left( 1-A_{\bar{\sigma}
}\left( \varepsilon \right) \right) +B_{\bar{\sigma}}^{<}\left( \varepsilon
\right) -A_{\bar{\sigma}}^{<}\left( \varepsilon \right) \left( G_{\sigma
}^{a}\left( \varepsilon \right) ^{-1}+\Sigma \left( \varepsilon \right)
\right) }{1-\left\langle n_{\bar{\sigma}}\right\rangle -A_{\bar{\sigma}
}\left( \varepsilon \right) }G_{\sigma }^{a}\left( \varepsilon \right) .
\label{G-2}
\end{equation}
\end{widetext}
Here the self-energy $\Sigma \left( \varepsilon \right) =\frac{-i\Gamma
\left( \varepsilon \right) -\Phi \left( \varepsilon \right) -\left( \Phi
\left( \varepsilon \right) \right) ^{\ast }}{1+\pi \Pi \left( \varepsilon
\right) },$ $\Gamma \left( \varepsilon \right) =\Gamma _{g}\left(
\varepsilon \right) +\Gamma _{t}\left( \varepsilon \right) $ and $\Sigma
^{<}\left( \varepsilon \right) =\frac{i\Xi _{1}\left( \varepsilon \right)
+\Omega \left( \varepsilon \right) }{1+\pi \Pi \left( \varepsilon \right) }$%
, the average of occupation $\left\langle n_{\bar{\sigma}}\right\rangle =-%
\frac{1}{\pi }\int \frac{\Xi _{1}\left( \varepsilon \right) }{\Gamma \left(
\varepsilon \right) }\text{Im}(G_{\bar{\sigma}}^{r}\left( \varepsilon \right))
d\varepsilon ,$ and
\begin{equation}
A_{\bar{\sigma}}\left( \varepsilon \right) =\frac{1}{\pi }\int \frac{g\left(
\varepsilon ,\varepsilon ^{\prime }\right) \left( \Xi _{1}\left( \varepsilon
^{\prime }\right) +i\Omega \left( \varepsilon ^{\prime }\right) \right) (G_{%
\bar{\sigma}}^{r}\left( \varepsilon ^{\prime }\right) )^{\ast }}{1+\pi \Pi
\left( \varepsilon ^{\prime }\right) }d\varepsilon ^{\prime },  \label{A}
\end{equation}%
\begin{eqnarray}
B_{\bar{\sigma}}\left( \varepsilon \right)  &=&\frac{1}{\pi }\int g\left(
\varepsilon ,\varepsilon ^{\prime }\right) \Xi _{1}\left( \varepsilon
^{\prime }\right) d\varepsilon ^{\prime }+  \notag \\
&&\frac{i}{\pi }\int g\left( \varepsilon ,\varepsilon ^{\prime }\right)
\left( \Xi _{2}\left( \varepsilon ^{\prime }\right) +\Xi _{3}\left(
\varepsilon ^{\prime }\right) \right) (G_{\bar{\sigma}}^{r}\left(
\varepsilon ^{\prime }\right) )^{\ast }d\varepsilon ^{\prime }  \notag \\
&&+\frac{1}{\pi }\int \frac{g\left( \varepsilon ,\varepsilon ^{\prime
}\right) \Xi _{1}\left( \varepsilon ^{\prime }\right) \Phi \left(
\varepsilon ^{\prime }\right) (G_{\bar{\sigma}}^{r}\left( \varepsilon
^{\prime }\right) )^{\ast }}{1+\pi \Pi \left( \varepsilon ^{\prime }\right)}
d\varepsilon ^{\prime },\label{B}
\end{eqnarray}%
\begin{equation}
A_{\bar{\sigma}}^{<}\left( \varepsilon \right) =\frac{i\Xi _{1}\left(
\varepsilon \right) \text{Re}\left( G_{\bar{\sigma}}^{r}\left( \varepsilon
\right) \right) +i\Omega \left( \varepsilon \right) \text{Im}\left( G_{\bar{%
\sigma}}^{r}\left( \varepsilon \right) \right) }{1+\pi \Pi \left(
\varepsilon \right) },  \label{AL}
\end{equation}%
\begin{eqnarray}
B_{\bar{\sigma}}^{<}\left( \varepsilon \right)  &=&i\Xi _{1}\left(
\varepsilon \right) +i\left( \Xi _{2}\left( \varepsilon \right) +\Xi
_{3}\left( \varepsilon \right) \right) \text{Im}\left( G_{\bar{\sigma}%
}^{r}\left( \varepsilon \right) \right)   \notag \\
&&-i\Xi _{1}\left( \varepsilon \right) \frac{\Phi \left( \varepsilon \right)
\text{Im}\left( G_{\bar{\sigma}}^{r}\left( \varepsilon \right) \right) }{%
1+\pi \Pi \left( \varepsilon \right) },  \label{BL}
\end{eqnarray}%
where $g\left( \varepsilon ,\varepsilon ^{\prime }\right) =1/\left(
\varepsilon -\varepsilon ^{\prime }+i0^{+}\right) $, $\Omega \left(
\varepsilon \right) =\Phi \left( \varepsilon \right) f_{g}\left( \varepsilon
\right) +\left( \Phi \left( \varepsilon \right) \right) ^{\ast }f_{t}\left(
\varepsilon \right) ,$ $\Xi _{1}\left( \varepsilon \right) =f_{g}\left(
\varepsilon \right) \Gamma _{g}\left( \varepsilon \right) +f_{t}\left(
\varepsilon \right) \Gamma _{t}\left( \varepsilon \right) ,$ $\Xi _{2}\left(
\varepsilon \right) =f_{g}\left( \varepsilon \right) \Gamma _{g}\left(
\varepsilon \right) \Gamma _{g}\left( \varepsilon \right) +f_{t}\left(
\varepsilon \right) \Gamma _{t}\left( \varepsilon \right) \Gamma _{t}\left(
\varepsilon \right) ,$ $\Xi _{3}\left( \varepsilon \right) =\Gamma
_{g}\left( \varepsilon \right) \Gamma _{t}\left( \varepsilon \right) \left(
f_{gt}\left( \varepsilon \right) +f_{tg}\left( \varepsilon \right) \right) $
with $f_{gt}\left( \varepsilon \right) =f_{g}\left( \varepsilon \right)
\left( 1-f_{t}\left( \varepsilon \right) \right) $. At the hand of the above
expressions, we can calculate numerically the Green's function $G_{\sigma
}^{r}\left( \varepsilon \right) $ and $G_{\sigma }^{a}\left( \varepsilon
\right) =\left( G_{\sigma }^{r}\left( \varepsilon \right) \right) ^{\ast }$
in a self-consistent way, and further the lesser Green function $G_{\sigma
}^{<}\left( \varepsilon \right) $. Finally, we obtain the current $I$ by Eq.
(\ref{I}).

\section{Numerical results and discussions}

\label{sec3}

In the following numerical calculation, several input parameters are tuned
in order to study the system in different physical situations. The atomic
level $\epsilon_d$ is changed to study the different regimes ranged from the
Kondo regime, to the mixed valence regime, even to the empty orbital regime.
\cite{ Goldhaber-Gordon1998} The chemical potential of graphene $\mu$ can be
tuned by gate-voltage, which makes the Fermi level is located at ($\mu = 0$%
), above ($\mu > 0$) or below ($\mu < 0$) the Dirac points. The different
chemical potential will lead to quite different physics, as shown below. The
voltage bias $V_{sd}$ is also applied between STM tip and graphene sheet to
calculate the differential conductance. For convenience, we take $\Gamma_0$
as the units of energy in the calculations. As a systematic study, we first
to calculate the Kondo effect of the adatom plus the graphene when the STM
tip is omitted. The aim is to explore the influence of the linear dispersion
around the Dirac points to the Kondo effect and compare it with the adatom on
normal metal surface. Subsequently, we consider the influence of the
STM tip and explore the Kondo resonance when both the normal metal and the
graphene are coupled to the adatom. Finally, we turn on the direct channel
between the STM tip and the Graphene and check its effect in the tunneling
spectroscopy, which is more closely related to the experiment.

\subsection{Kondo effect of the adatom without the STM tip}

\begin{figure}[t]
\center{\includegraphics[clip=true,width=\columnwidth]{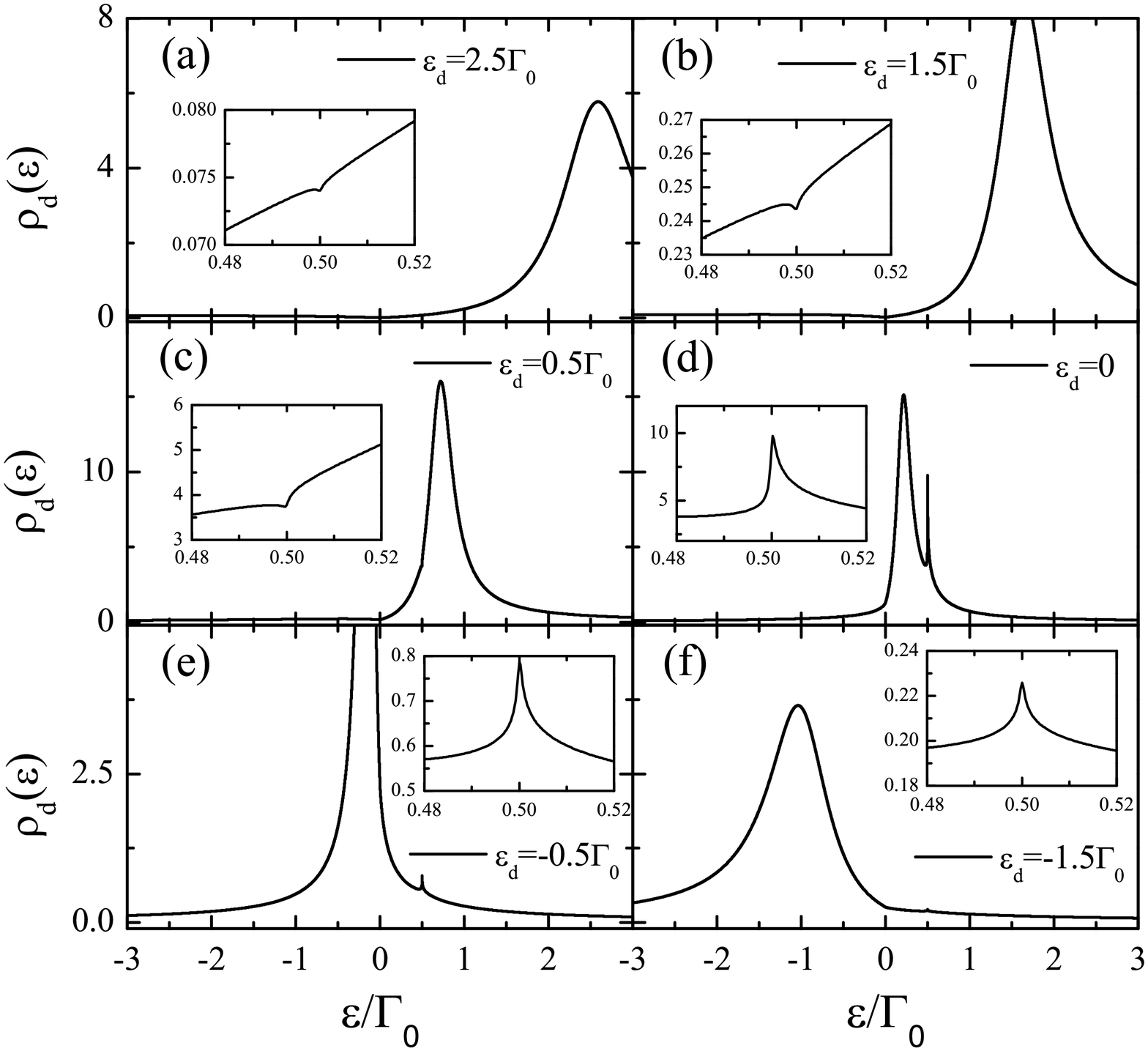}}
\caption{The Kondo resonance of the magnetic adatom on graphene with
different atomic levels ranged from the empty orbital regime to the Kondo
regime. From (a) to (f) the atomic level is $\protect\epsilon %
_{d}=2.5\Gamma_{0}$, $1.5\Gamma_{0}$, $0.5\Gamma_{0}$, 0, $-0.5\Gamma_{0}$, $-1.5\Gamma_{0}$.
The inset in each panel is enlarged view around the Kondo
resonance. The chemical potential is $\protect\mu=0.5\Gamma_{0}$, the temperature $T=0.00005\Gamma_{0}$,
and the half-bandwidth $D=5\Gamma_{0}$.}
\label{fig2}
\end{figure}

\begin{figure}[t]
\center{\includegraphics[clip=true,width=\columnwidth]{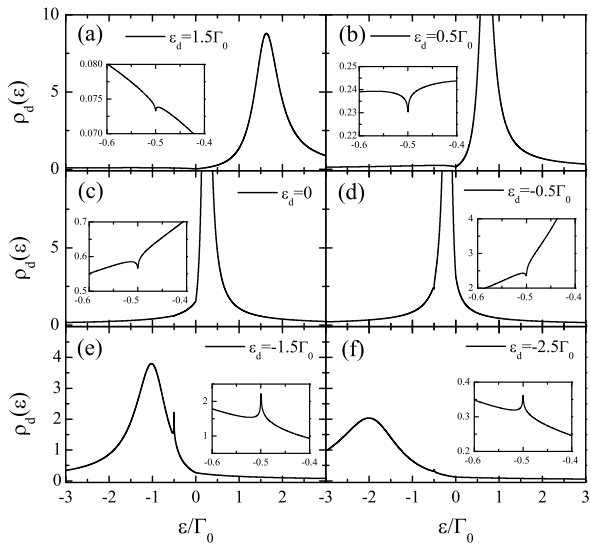}}
\caption{The same as Fig. \protect\ref{fig2} but for $\protect\mu =
-0.5\Gamma_0 $. From (a) to (f) the atomic level is $\protect\varepsilon %
_{d}=1.5\Gamma_{0}$, $0.5\Gamma_{0}$, 0, $-0.5\Gamma_{0}$, $-1.5\Gamma_{0}$, $-2.5\Gamma_{0}$.}
\label{fig3}
\end{figure}

\begin{figure}[t]
\center{\includegraphics[clip=true,width=0.8\columnwidth]{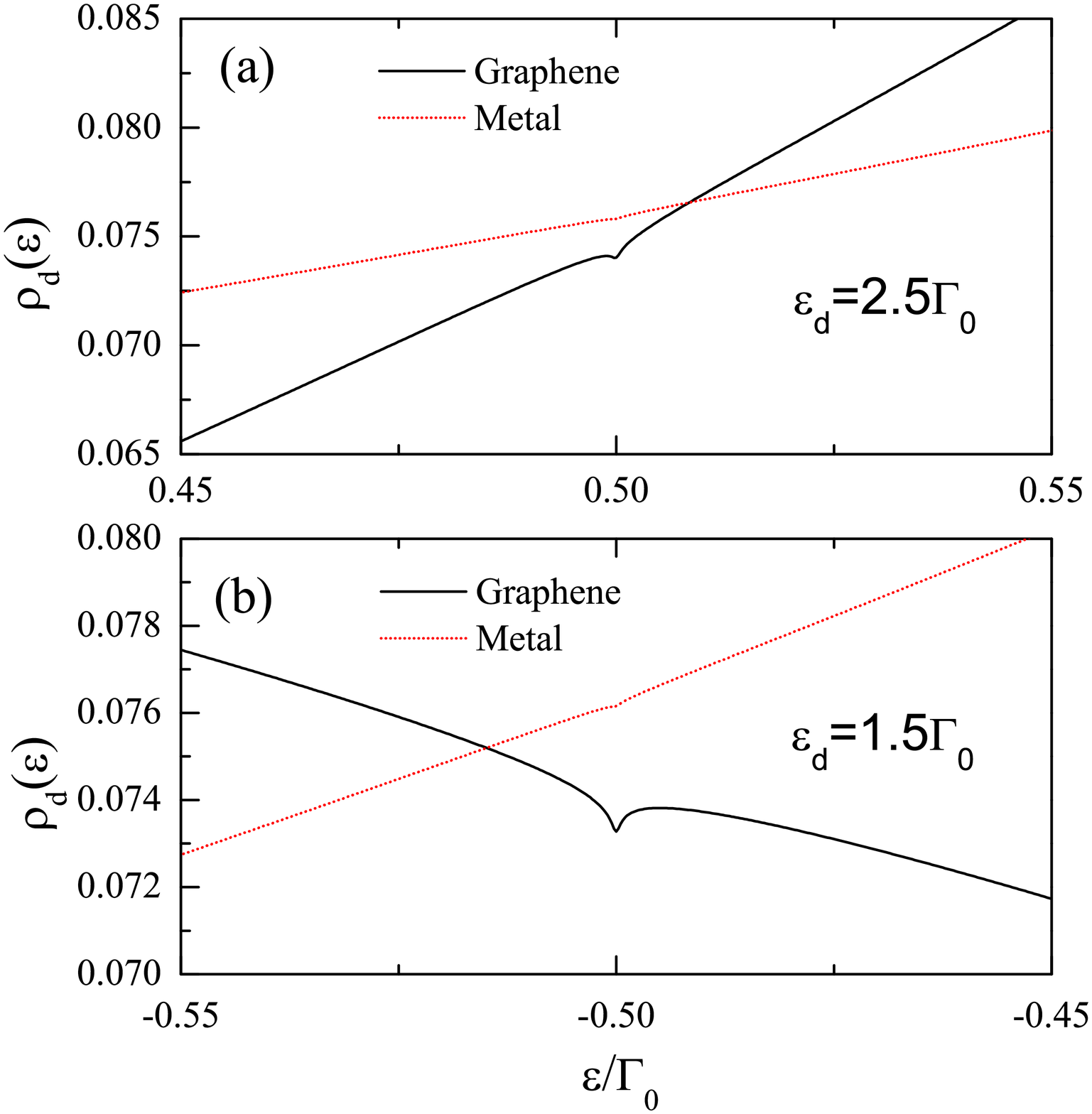}}
\caption{Comparison between the density of states around the Fermi level in
the empty orbital regime for the metallic (red dashed lines) and the
graphene (solid lines) hosts. (a) $\protect\mu = 0.5\Gamma_0$, $\protect\varepsilon_d = 2.5\Gamma_0$,
and (b) $\protect\mu = -0.5\Gamma_0$, $\protect\varepsilon_d = 1.5\Gamma_0$.}
\label{fig4}
\end{figure}

In order to explore the effect of the linear dispersion around the Dirac
points, we first the system composed of the magnetic adatom plus the
graphene and omit the STM tip. As pointed out by Uchoa \textit{et al.}, \cite%
{Uchoa2008} it is more easier to form the local moment for the magnetic
impurity on the surface of the graphene than in the normal metal host. This
conclusion has a direct consequence that one can observe the Kondo resonance
in a more wider parameter region for the former than the latter one. This is
really true as shown in Figs. \ref{fig2} and \ref{fig3} where we show the
local density of states of the adatom given by $\rho_d(\varepsilon) = -\frac{%
1}{\pi}\text{Im}G^r_{\sigma}(\varepsilon)$. In Fig. \ref{fig2} we consider $\mu = 0.5\Gamma_0$,
namely, the Fermi level is above the Dirac points. We change the atomic
level from the empty orbital regime, as shown in Fig. \ref{fig2}(a) to the
Kondo regime [see Fig. \ref{fig2}(f)], around the Fermi level the local
density of states show special feature, i.e., the Kondo resonance, although
their shapes are quite different. In the Kondo regimes, the Kondo resonance
exhibits as a sharp peak around the Fermi level but one should notice that
with increasing the atom level the peak becomes gradually asymmetric, which
is quite obvious in Fig. \ref{fig2}(d) for $\varepsilon_d = 0$.
Continuing to change the atomic level to $\varepsilon_d = 0.5\Gamma_{0}$, the Kondo
resonance has a typical peak-dip structure. Finally, we move the atomic
level to the empty orbital regimes, the Kondo effect manifests as an
anti-resonance, which is clearly seen from the inset of Fig. \ref{fig2}(a)
and (b). These features are nothing but the Kondo resonance. They can be
understood based on our previous work. \cite{Luo2004} In such a system, the
Kondo resonance results from the coherent supposition between the Kondo peak
and the broadening of the impurity level due to the hybridization with the
conduction electron. In the Kondo regime, the broadening of the impurity
level is quite weak around the Fermi level so the Kondo resonance shows as a
sharp peak. When the system approaches to the mixed valence regime, the
broadening of the impurity level becomes more and more sizeable and as a
result, the Kondo peak distorts gradually due to the interference, and
finally it becomes an anti-resonance when the system is in the empty orbital
regime since in this case the Fermi level is located at the left-hand side
of the broadening impurity level. The essential reason that the Kondo effect can exist in a wide parameter
region is the impurity level is broadened extensively due to the linear
density of states of the graphene around the Dirac points. Therefore, the
adatom on the surface of the graphene provides an idea system to study the
Kondo effect in various parameter spaces. The same physics has been observed
when the Fermi level is below the Dirac points, as shown in Fig. \ref{fig3}(a)-(f).

To confirm above observation, we compare the local
density of states around the Fermi level for the adatom on graphene and on
metal surface in the empty orbital regime, as show in Fig. \ref{fig4}. For metal host, one cannot observe any feature, which is in contrast to that in the graphene, where an obvious anti-resonance exists. The results are true either the chemical potential is above or below the Dirac points.

\subsection{Kondo Effect of the adatom with the STM tip}

\begin{figure}[tbp]
\center{\includegraphics[clip=true,width=\columnwidth]{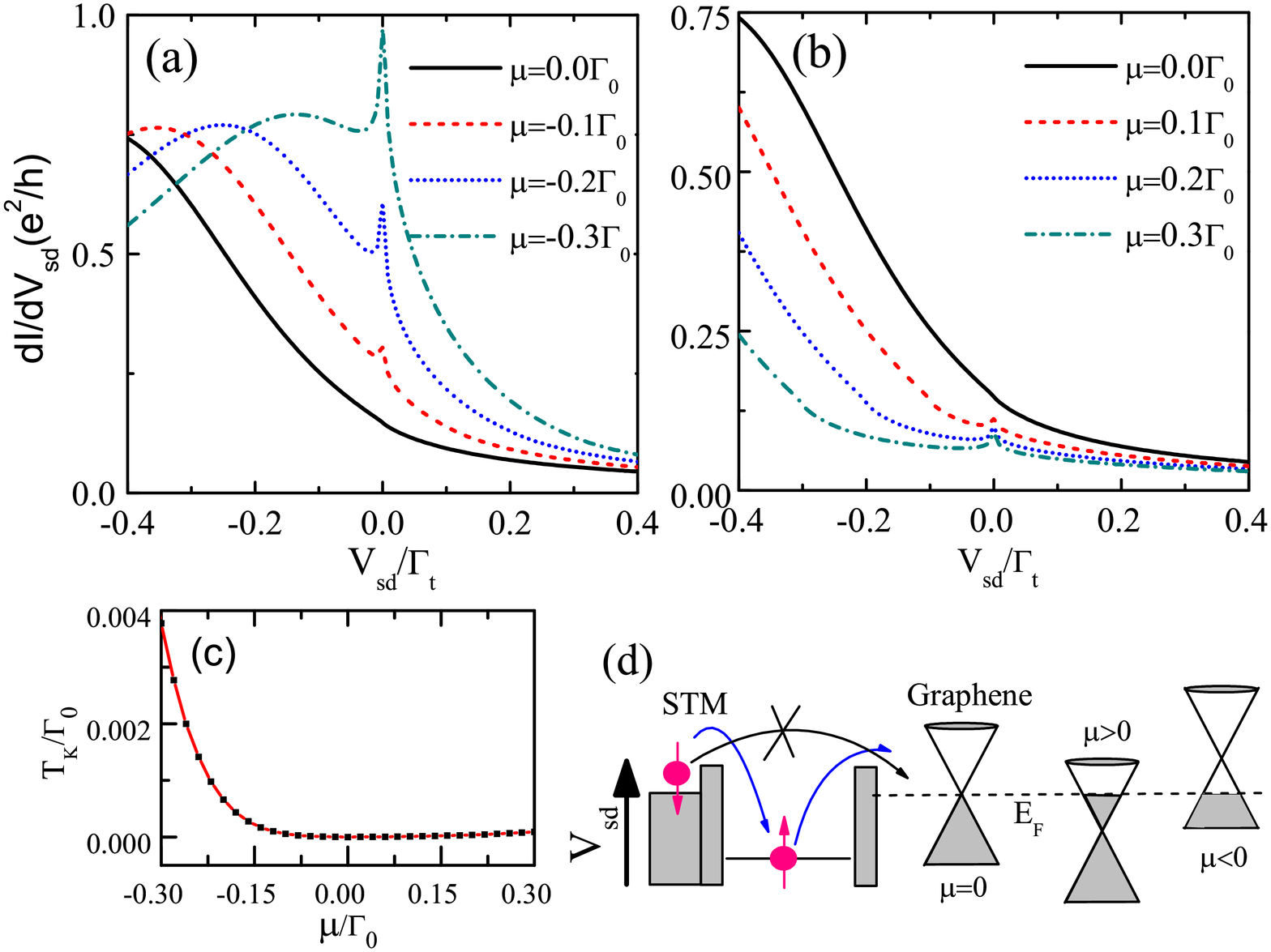}}
\caption{The Kondo resonance around the zero bias varies with the chemical
potential (a) $\protect\mu < 0$ and (b) $\protect\mu > 0$. (c) The Kondo
temperature $T_k$ versus the chemical potential $\protect\mu$. The other
parameters used are $\protect\epsilon_d = -\Gamma_0$ where $\Gamma_t =
0.1 \Gamma_0$, $T_{tg} = 0$ and $D = 5 \Gamma_0$. (d) shows
schematically the cotunneling processes with the chemical potential $\protect\mu < 0$
and $\protect\mu > 0$, respectively.}
\label{fig5}
\end{figure}

When the STM tip is considered, the adatom couples with two baths, one is
the STM tip which is a normal metal with a constant density of states, as
studied in the conventional Kondo physics. The other is the graphene which
has a linear density of states around the Dirac points. We first study the
nonequilibrium Kondo effect in this case. Therefore, we omit the direct channel
between the STM tip and the graphene. In the calculations, the coupling
between the adatom and tip is taken with $\Gamma_t = 0.1 \Gamma_0$.
Fig. \ref{fig5} shows the differential conductance around the zero bias. When the chemical potential $\mu$ is tuned to meet the Dirac points, the curve at the zero bias shows a kink but no visible Kondo peak, which is a consequence of the zero density of states at
the Dirac points. When the chemical potential $\mu$ is tuned to be away from
the Dirac points, either above or below the Dirac points, the zero bias peak
occurs, which shows the Kondo effect. When the chemical potential below the Dirac points,
the more larger $|\mu|$ is, the more wider the peak is. This is easy to understand
since the larger $\mu$ means the larger density of states at the side of graphene.
When $\mu$ is above the Dirac points, the width of the
Kondo resonance keeps almost unchange (see Fig. \ref{fig5}(c)), which also can be obtained from the chemical potential dependent Kondo temperature.\cite{Zhu2011}
However, when $\mu$ is below the Dirac points, the Kondo temperature $T_K$ shows obvious dependence on $\mu$. As a result, the Kondo temperature is obvious asymmetric, which is consistent with that reported in the literature. \cite{Vojta2010,Zhu2011}
This is due to the fact of the particle-hole asymmetry in graphene.
In addition, one also notes that in both cases the curves show a kink when
the bias scans through the Dirac points.

\subsection{Scanning tunneling spectroscopy with direct channel}

\begin{figure}[tbp]
\includegraphics[clip=true,width=0.8\columnwidth]{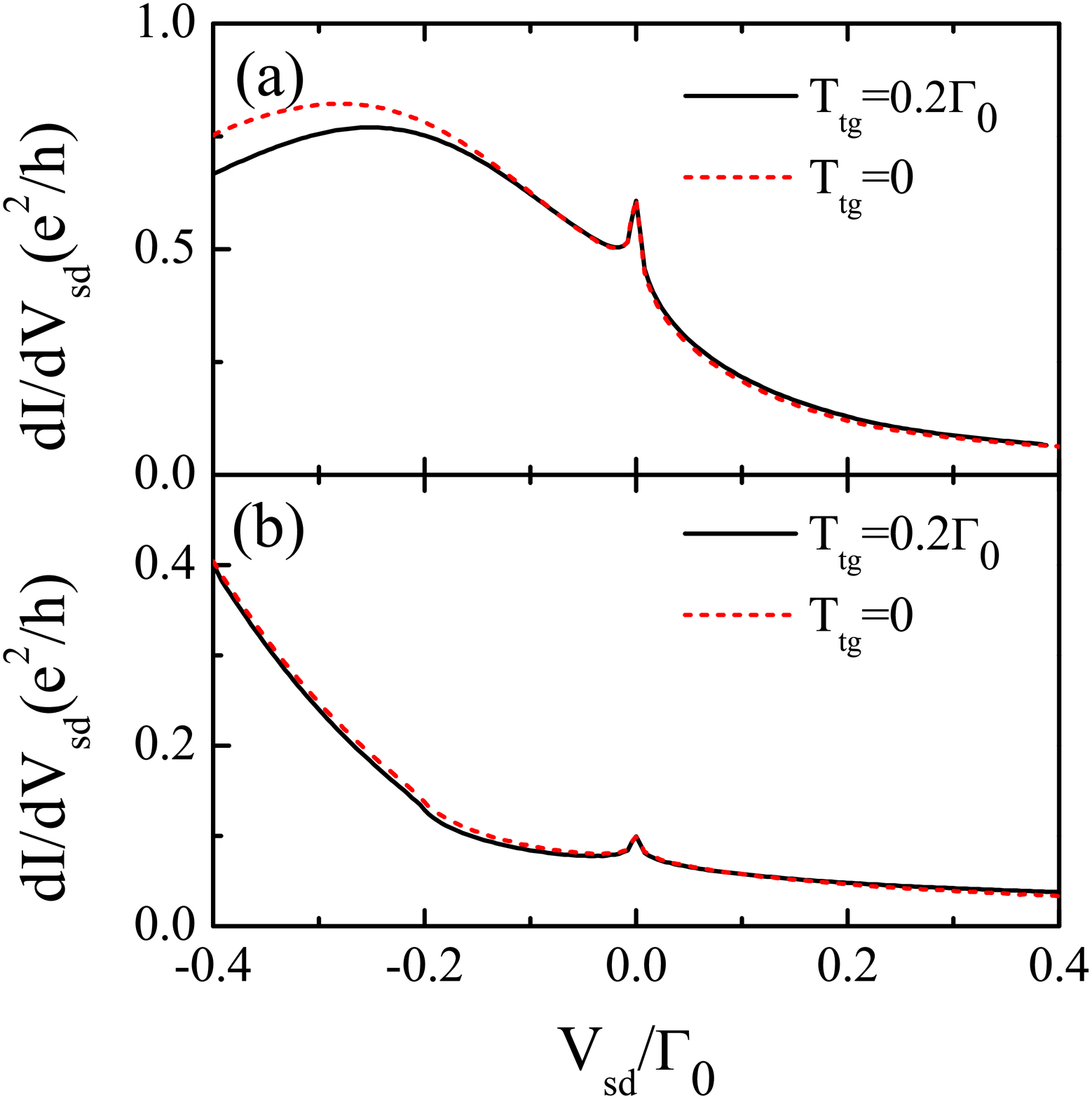}
\caption{The Kondo resonance around the zero bias with and without the
direct tunneling $T_{tg}$ channel for two cases (a) $\protect\mu =
-0.2\Gamma_0$ and (b) $\protect\mu = 0.2\Gamma_0$. We take $\protect%
\varepsilon_d = -\Gamma_0$ and $T_{tg} = 0.2\Gamma_0$. The other
parameters used are the same as Fig. \protect\ref{fig5}.}
\label{fig6}
\end{figure}

In this subsection we discuss the scanning tunneling spectroscopy of the
adatom on the surface of graphene by switching on the direct tunneling
channel between the STM tip and graphene, a more realistic case in
experiments. Due to the additional direct channel, one may expect that the
STM spectroscopy should be changed dramatically leaded by the interference
between the direct and the indirect channels, as usually observed in a system with the adatom on the surface of the normal
metals. \cite{Madhavan1998} However, the result is somehow surprising, as
shown in Fig. \ref{fig6}. In comparison to the spectroscopy without the
direct channel, except for the amplitudes, the shape almost keeps unchanged
for both $\mu > 0$ and $\mu < 0$. To check the underlying physics of this
result, let us recall the Fano resonance theory. \cite{Fano1961} In the STM
case, due to the adatom scattering and its interference with the direction
channel, the differential conductance of the conduction electron can be
written as \cite{Luo2004}
\begin{equation}  \label{fano}
\frac{dI}{dV} \propto (q^2 -1) \text{Im}\, G(\varepsilon) - 2 q \text{Re}\,
G(\varepsilon),
\end{equation}
where $G(\varepsilon)$ denotes the adatom Green's function and $q \approx -%
\text{Re}\, g(\varepsilon)/\text{Im}\, g(\varepsilon)$ is the Fano asymmetry
factor. Here $g(\varepsilon)$ is the graphene Green's function with $\text{Re%
}\, g(\varepsilon) = (\varepsilon/D^2)\ln|\varepsilon^2/(D^2 -
\varepsilon^2)|$ and $\text{Im}\, g(\varepsilon) = -\pi
|\varepsilon|/D^2\Theta(D-|\varepsilon|)$. Thus the $q$ factor is $2\,\text{%
sgn}(\mu)/\pi \ln |\mu|/D$ near the Fermi level. \cite{Wehling2010b} For
graphene, $D\sim 6\text{eV}$ (Ref. [\onlinecite{Wehling2007}]) and if $|\mu|$
maximally is in the order of $0.1\text{eV}$, so $|q| \gtrsim 6$. That means
that the first term in Eq. (\ref{fano}) dominates the second term, as a
result, the $dI/dV$ has the almost same profile as $\text{Im}G(\varepsilon)$, the local density of states of the adatom. This is the reason that the
shapes of the Kondo resonance around the zero bias is almost same despite
the direct channel is switched on or off.

\section{Summary}

\label{sec4}

In summary, we study systematically the Kondo effect of a magnetic adatom on
the surface of graphene and its STM spectra. The main conclusion is that the
Kondo resonance can exist in a much wide parameter region ranged from the
Kondo regime, to the mixed valence regime, even to the empty orbital regime.
In the mixed valence or the empty orbital regimes the Kondo resonance show specially
lineshape due to the Fano resonance since in these regimes the broadening of
the impurity level provides a significant background. The STM spectra have a
particle-hole asymmetry when the chemical potential in graphene is tuned by
the gate voltages. Furthermore, it is found that the direct tunneling
channel between the STM tip and graphene has no obvious effect on the
lineshape of the Kondo resonance around the zero bias, which is a
consequence of the unusual large asymmetry factor in graphene host. The rich
behaviors of the Kondo resonance obtained indicate that the graphene is an
ideal platform to study the Kondo physics and these results are useful to
further stimulate the relevant experimental works in this system.

\section{Acknowledgement}

Supported from CMMM of Lanzhou University, the NSF, the national program for
basic research and the fundamental research funds for the central
universities of China is acknowledged.


\begin{thebibliography}{99}
\bibitem{Novoselov2004} K. S. Novoselov, A. K. Geim, S. V. Morozov, D.
Jiang, Y. Zhang, S. V. Dubonos, I. V. Grigorieva, and A. A. Firsov, Science
\textbf{306}, 666 (2004).

\bibitem{Novoselov2005} K. S. Novoselov, A. K. Geim, S. V. Morozov, D.
Jiang, M. I. Katsnelson, I. V. Grigorieva, S. V. Dubonos, and A. A. Firsov,
Nature \textbf{438}, 197 (2005).

\bibitem{Beenakker2008} Beenakker, C. W. J. Colloquium: Andreev reflection
and Klein tunneling in graphene. Rev. Mod. Phys. \textbf{80}, 1337 (2008).

\bibitem{Neto2009} A. H. Castro Neto, F. Guinea, N. M. R. Peres, K. S.
Novoselov, and A. K. Geim, Rev. Mod. Phys. \textbf{81}, 109 (2009).

\bibitem{Peres2010} N. M. R. Peres, Rev. Mod. Phys. \textbf{82}, 2673 (2010).

\bibitem{Peres2006} N. M. R. Peres, F. Guinea, and A. H. Castro Neto, Phys.
Rev. B \textbf{73}, 125411 (2006).

\bibitem{Ziegler2006} K. Ziegler, Phys. Rev. Lett. \textbf{97}, 266802
(2006).

\bibitem{Nomura2007} K. Nomura and A. H. MacDonald, Phys. Rev. Lett. \textbf{%
98}, 076602 (2007).

\bibitem{Zhuang2009} H.-B. Zhuang, Q.-f Sun, and X. C. Xie, EPL \textbf{86},
58004 (2009).

\bibitem{Kotov2010} V. N. Kotov, B. Uchoa, V. M. Pereira, A. H. C. Neto, and
F. Guinea, arxiv:1012.3484 (2010).

\bibitem{Jacob2010} D. Jacob and G. Kotliar, Phys. Rev. B \textbf{82},
085423 (2010).

\bibitem{Anderson1961} P. W. Anderson, Phys. Rev. \textbf{124}, 41 (1961).

\bibitem{Uchoa2008} B. Uchoa, V. N. Kotov, N. M. R. Peres, and A. H. Castro
Neto, Phys. Rev. Lett. \textbf{101}, 026805 (2008).

\bibitem{Cornaglia2009} P. S. Cornaglia, G. Usaj, and C. A. Balseiro, Phys.
Rev. Lett. \textbf{102}, 046801 (2009).

\bibitem{Hu2011} F. M. Hu, Tianxing Ma, Hai-Qing Lin, and J. E. Gubernatis,
Phys. Rev. B \textbf{84}, 075414 (2011).

\bibitem{Sengupta2008} K. Sengupta and G. Baskaran, Phys. Rev. B \textbf{77}%
, 045417 (2008).

\bibitem{Goldhaber-Gordon1998} D. Goldhaber-Gordon, J. G\"ores, M. A.
Kastner, H. Shtrikman, D. Mahalu, and U. Meirav, Phys. Rev. Lett. \textbf{81}%
, 5225 (1998).

\bibitem{Uchoa2009} B. Uchoa, L. Yang, S. W. Tsai, N. M. R. Peres, and A. H.
Castro Neto, Phys. Rev. Lett. \textbf{103}, 206804 (2009).

\bibitem{Saha2010} K. Saha, I. Paul, and K. Sengupta, Phys. Rev. B \textbf{81%
}, 165446 (2010).

\bibitem{Wehling2010a} T. O. Wehling, A. V. Balatsky, M. I. Katsnelson, A.
I. Lichtenstein, and A. Rosch, Phys. Rev. B \textbf{81}, 115427 (2010).

\bibitem{Wehling2010b} T. O. Wehling, H. P. Dahal, A. I. Lichtenstein, M. I.
Katsnelson, H. C. Manoharan, and A. V. Balatsky, Phys. Rev. B \textbf{81},
085413 (2010).

\bibitem{Uchoa2011} B. Uchoa, T. G. Rappoport, and A. H. Castro Neto, Phys.
Rev. Lett. \textbf{106}, 016801 (2011).

\bibitem{Vojta2010} M. Vojta, L. Fritz, and R. Bulla, EPL \textbf{90}, 27006
(2010).

\bibitem{Fano1961} U. Fano, Phys. Rev. \textbf{124}, 1866 (1961).

\bibitem{Luo2004} H. G. Luo, T. Xiang, X. Q. Wang, Z. B. Su, and L. Yu,
Phys. Rev. Lett. \textbf{92}, 256602 (2004); \textbf{96}, 019702 (2006).


\bibitem{Zhu2011} Z.-G. Zhu, and J. Berakdar, Phys. Rev. B \textbf{84},
165105 (2011).

\bibitem{Jauho1994} A. P. Jauho, N. S. Wingreen and Y. Meir, Phys. Rev. B \textbf{50}, 5528 (1994).

\bibitem{Mahan1993} G. D. Mahan, \textit{Many-Particle Systems}, 2nd ed.
(Plenum Press, New York, 1993).

\bibitem{Niu1999} C. Niu, D.L. Lin, and T.H. Lin, J. Phys.: Condens. Matter
\textbf{11}, 1511 (1999).


\bibitem{Swirkowicz2003} R. \'{S}wirkowicz, J. Barna\'{s}, and M. Wilczy\'{n}%
ski, Phys. Rev. B \textbf{68}, 195318 (2003).

\bibitem{Krawiec2004} M. Krawiec and K. I. Wysoki\'{n}ski, Supercond. Sci.
Technol. \textbf{17}, 103-112 (2004).

\bibitem{Zubarev1960} D. N. Zubarev, Usp. Fiz. Nauk \textbf{71}, 71 (1960)
[Sov. Phys. Usp. \textbf{3}, 320 (1960)].

\bibitem{Lacroix1981} C. Lacroix, J. Phys. F \textbf{11}, 2389 (1981).

\bibitem{Luo1999} H.-G. Luo, Z.-J. Ying and S.-J. Wang, Phys. Rev. B \textbf{%
59}, 9710 (1999).

\bibitem{Madhavan1998} V. Madhavan, W. Chen, T. Jamneala, M. F. Crommie, and
N. S. Wingreen, Science \textbf{280}, 567 (1998).

\bibitem{Wehling2007} T. O. Wehling, A. V. Balatsky, M. I. Katsnelson, A. I.
Lichtenstein, K. Scharnberg, and R. Wiesendanger, Phys. Rev. B \textbf{75},
125425 (2007).

\end{thebibliography}
\end{document}